\documentclass[11pt]{article}

\usepackage{fullpage,latexsym,amsmath,amsfonts,color,epsfig}
\usepackage[latin1]{inputenc} 

\newcommand{\barH}{{\bar H}}

\newcommand{\bfT}{{\bf T}}

\newcommand{\calI}{{\cal I}}
\newcommand{\calJ}{{\cal J}}

\newcommand{\calX}{{\cal X}}
\newcommand{\calY}{{\cal Y}}

\newcommand{\pmtn}{{\mbox{\rm pmtn}}}

\newcommand{\onehalf}{{\mbox{$\frac{1}{2}$}}}

\newcommand{\onefourth}{{\mbox{$\frac{1}{4}$}}}

\newcommand{\braced}[1]{{ \left\{ #1 \right\} }}

\DeclareMathSymbol{\reals}{\mathbin}{AMSb}{"52}

\newcommand{\calXbef}[1]{\calX_{<#1}}
\newcommand{\calXaft}[1]{\calX_{>#1}}

\newcommand{\threepart}{{\mbox{\sc 3-Partition}}}
\newcommand{\meanflowsched}{{\mbox{$P|r_j,\mbox{\rm pmtn}|\sum C_j$}}}

\newcommand{\undefined}{{\bot}}

\newtheorem{theorem}{Theorem}

\newtheorem{lemma}[theorem]{Lemma}

\newtheorem{claim}[theorem]{Claim}

\newenvironment{proof}{{\it Proof:\/}}{$\Box$\vskip 0.1in}

                {$\spadesuit$\smallskip}

\newenvironment{bigeqn*}{\large\begin{eqnarray*}}{\end{eqnarray*}}

\newcommand{\SetFigFont}[6]{#6}

\begin{document}


\title{The Complexity of Mean Flow Time Scheduling Problems\\
                 with Release Times}

\author{
Philippe Baptiste%
\thanks{CNRS LIX, Ecole Polytechnique,
91128 Palaiseau, France.
\{baptiste,durr\}@lix.polytechnique.fr.
Supported by the NSF/CNRS grant 17171 and ANR/Alpage. }
\and
Peter Brucker%
\thanks{Universit\"at Osnabr\"uck, 
Fachbereich Mathematik/Informatik,
49069 Osnabr\"uck, Germany. 
peter@mathematik.uni-osnabrueck.de.
Supported by INTAS Project 00-217 and by DAAD PROCOPE Project D/0427360.}
\and
Marek Chrobak%
\thanks{Department of Computer Science,
University of California,
Riverside, CA 92521.
marek@cs.ucr.edu.
Supported by NSF grants CCR-0208856 and INT-0340752.
}
\and
Christoph Dürr%
\footnotemark[1]
\and
Svetlana A. Kravchenko%
\thanks{United Institute of Informatics Problems,
Surganova St. 6, 220012 Minsk, Belarus.
kravch@newman.bas-net.by.
Supported by the Alexander von Humboldt Foundation.
}
\and
Francis Sourd%
\thanks{CNRS LIP6,
Universit\'e Pierre et Marie Curie,
Place de Jussieu, 75005 Paris, France.
Francis.Sourd@lip6.fr.}
}

\maketitle

\begin{abstract}
We study the problem of preemptive scheduling $n$ 
jobs with given release times on $m$ identical parallel machines.
The objective is to minimize the average flow time. 
We show that when all jobs have equal processing times then
the problem can be solved in polynomial time using linear programming.
Our algorithm can also be applied to the open-shop problem with release
times and unit processing times.
For the general case (when processing times are arbitrary), we
show that the problem is unary NP-hard.
\end{abstract}

\section{Introduction}

In the scheduling problem we study, the input instance consists of $n$
jobs with given release times and processing times.
The objective is to compute a preemptive schedule of those
jobs on $m$ machines that minimizes the average flow time or,
equivalently, the sum of completion times, $\sum C_j$. In the standard
scheduling notation, the problem can be described as
$P|r_j,\pmtn|\sum C_j$.

First, we focus on the case when all jobs have the same
processing time $p$, that is $P|r_j, p_j = p,\pmtn|\sum C_j$.
Herrbach and Leung \cite{HL90} showed
that, for $m=2$, the optimal schedule can be computed in time
$O(n \log n)$. 
For more machines, the complexity of the problem was open.
Addressing this open problem, we present an
algorithm whose running time is polynomial in $m$ and $n$.

Our algorithm is based on linear programming.
We show that there is always an optimal schedule in a certain
\emph{normal} form. We then give a
simple linear program of size $O(mn)$, which directly
defines an optimal normal schedule. 
Since the coefficients in the constraints of our
linear program are $-1$, $0$, or $1$, the result of
Tardos \cite{Tardos86} implies that the problem can be solved
in worst case time $O(n^{5}m^{5})$.

We show that, without loss of generality, we can assume
that the preemptions occur only at integer times. This
yields a polynomial-time algorithm for the open shop
problem $O | r_i, p_{ij}=1 | \sum C_i$, for it is
known that this problem is equivalent to
$P | r_i, p_i=m, \mbox{pmtn}^+ | \sum C_i,$ where $m$ 
is the number of machines \cite{brucker.jurisch.ea:open-shop}.  
(Notation $\mbox{pmtn}^+$ means that preemptions are allowed
only at integer times.) Previously, it was only known that
this problem can be solved in polynomial time if $m$
is constant \cite{Tautenhahn.Woeginger:Minimizing-the-total}.

In the open shop problem $O | r_i, p_{ij}=1 | \sum C_i$
each job has to be processed on each machine exactly once and with a unit processing time. At any time each machine can execute at most one job and every job can be scheduled by at most one machine. No job can start before its given release time, and the goal is to minimize the total completion time.

In the last section we consider the general case, when the
processing times are arbitrary.
Du, Leung and Young \cite{DLY90} proved
this problem is binary NP-hard for two machines.
We show that if the number of machines is not fixed then
the problem is in fact unary NP-complete.

We summarize the results discussed above in
Table~\ref{tab: complexity}.

\begin{table}[htb]
\begin{center}
\begin{tabular}{l|l}
        Problem & Complexity \\ \hline
  $P2 | r_j; \pmtn; p_j=p | \sum C_j$       
  & solvable in time $O(n\log n)$ \cite{HL90}    \\
  $P\phantom2 | r_j; \pmtn; p_j=p | \sum C_j$       
  & solvable in polynomial time [this paper]  \\
  $P2 | r_j; \pmtn\,\phantom{\;p_j=p} | \sum C_j$       
  &  binary NP-hard \cite{DLY90}            \\
  $P\phantom2 | r_j; \pmtn\,\phantom{\;p_j=p} | \sum C_j$       
  & unary NP-complete [this paper] \\
  $P\phantom2 | \phantom{r_j;}\,\pmtn ; p_j=p | \sum C_j$       
  & solvable by the greedy algorithm (trivial)  \\
  $P\phantom2 | r_j; \phantom{\pmtn\;}\,p_j=p | \sum C_j$       
  & solvable by the greedy algorithm (trivial)  \\
  $1\phantom{P} | r_j; \pmtn;  p_j=p | \sum w_j C_j$       
  & open  \\
  $P\phantom2 | r_j; \pmtn; p_j=p | \sum w_j C_j$       
  & unary NP-complete \cite{LY90}  \\
  $O\phantom{2} | r_j; \phantom{\pmtn\;}\,p_{ji}=1 | \sum C_j$
  & solvable in $O(n^{2}m^{6m})$ \cite{Tautenhahn.Woeginger:Minimizing-the-total},  in polynomial time [this paper] 
\end{tabular}
\end{center}
\caption{Complexity of related scheduling problems.
$P2$ stands for the two-machine problem, and $O$ for the open-shop problem. 
In problems with the
objective function $\sum w_j C_j$, each job $j$ is assigned a weight
$w_j\ge 0$, and the goal is to minimize the weighted sum of all completion
times.}
\label{tab: complexity}
\end{table}

\section{Structural Properties}
\label{sec: preemptions}

\paragraph{Basic definitions.}
Throughout the paper, $n$ and $m$ denote, respectively, the number of
jobs and the number of machines. The jobs are numbered $1,2,\dots,n$
and the machines are numbered $1,2,\dots,m$.  All jobs have the same
length $p$.  For each job $j$, $r_j$ is the release time of $j$,
where, without loss of generality, we assume that $0 = r_1 \le \ldots
\le r_n$. In this section we assume that all numbers 
$p,r_1,\ldots,r_n$ are integers; in the appendix
we show that our results can be extended to arbitrary real numbers.

We define a \emph{schedule} $\calX$ to be a function which, for any
time $t$, determines the set $\calX(t)$ of jobs that are running at
time $t$. This set $\calX(t)$ is called the \emph{profile} at time
$t$. Let $\calX^{-1}(j)$ denote the set of times when $j$ is executed,
that is $\calX^{-1}(j) = \braced{t: j\in \calX(t)}$. In addition we
require that $\calX$ satisfies the following conditions:
\begin{description}
\item{(s1)} At most $m$ jobs are executed at any time, that is
  $|\calX(t)|\le m$ for all times $t$.
\item{(s2)} No job is executed before its release time, that is, for
  each job $j$, if $t < r_j$ then $j\notin \calX(t)$.
\item{(s3)} Each job runs in a finite number of time intervals.  More
  specifically, for each job $j$, $\calX^{-1}(j)$ is a finite union of
  intervals of type $[s,t)$.
\item{(s4)} Each job is executed for time $p$, that is
  $|\calX^{-1}(j)| = p$.
\end{description}
In (s4), for a set $X$ of real numbers we use $|X|$ to denote its
measure.
It is not difficult to see that condition (s3) can be relaxed to allow
jobs to be executed in infinitely (but countably) many intervals,
without changing the value of the objective function.

By $C_j = \sup\calX^{-1}(j)$ we denote the completion time of a job
$j$.  In this paper, we are interested in computing a schedule that
minimizes the objective function $\sum_{j=1}^n C_j$.

\smallskip

Note that, since we are dealing with preemptive schedules, it does not
matter to which specific machines the jobs in $\calX(t)$ are assigned
to.  When such an assignment is needed, we will use the convention
that the jobs are assigned to machines in the increasing order of
indices (or, equivalently, release times): the job with minimum index
is assigned to machine $1$, the second smallest job to machine $2$,
etc.


\paragraph{Integral schedules.}
We say that a schedule $\calX$ is \emph{integral}, if $\calX(t)$ is constant for 
$t\in[s,s+1)$ at any $s\in\mathbb N$. (In other words, all preemptions occur
at integer times.) The following lemma is due to
\cite{Baptiste:Polynomial-time}, and we
include it here for the sake of completeness.


\begin{lemma} 
There exists an optimal schedule $\calX$ that is integral.
\end{lemma}

\begin{proof}
Let $\calX$ be an arbitrary optimal schedule. We
will show that there is an integral schedule $\calX'$ whose
objective value is not greater than that of $\calX$.
The proof is based on a flow network model of the scheduling problem,
and uses the fact that if all capacities are integral then there is an integral
solution.

Let $C_1,\ldots,C_n$ be the completion times in $\calX$.
We consider the set of all time points
\[
\{r_1,\ldots,r_n,\lfloor C_1\rfloor,
        \lceil C_1\rceil,\ldots,\lfloor C_n\rfloor,\lceil C_n\rceil\}.
\]
Suppose that there are $k$ distinct numbers in this set. 
We rename these numbers $t_1,...,t_k$ and order them in
increasing order, $t_1 < \ldots < t_k$.
These numbers define $k-1$ intervals $[t_i,t_{i+1})$ for $1\le i<k$.

We define a network which consists of the nodes $u_1,\ldots,u_n,v_1,\ldots,v_{k-1}$,
plus two more nodes designated as the source and the sink.
A node $u_j$ represents job $j$ and
a node $v_i$ represents interval $[t_i,t_{i+1})$.
For every job $j$, there is an arc from the source to $u_j$ with
capacity $p$ and cost $0$. 
For every time interval $[t_i,t_{i+1})$ there is an arc from $v_i$ to the sink
with capacity $m(t_{i+1}-t_i)$ and cost $0$.
For every job $j$ and every interval $[t_i,t_{i+1})\subseteq  [r_j,\lfloor C_j\rfloor )$,
there is an arc from $u_j$ to $v_i$ with capacity $t_{i+1}-t_i$ and cost $0$. 
In addition, for every job $j$ for which $C_j$ is not integral, there is an arc from
$u_j$ to $v_i$ for $t_i=\lfloor C_j\rfloor$ with capacity 
$1$ and cost $1$. (These are the only arcs with non zero cost.)

The schedule $\calX$ corresponds to a flow of value $np$, where the flow
on an arc $(u_j,v_i)$ has value $|\calX^{-1}(j)\cap [t_i,t_{i+1})|$.
The flows on other arcs are uniquely determined by the flows on all arcs
$[u_j,v_i)$. The cost of this flow is
$w = \sum_{j=1}^n (C_j- \lfloor C_j \rfloor )$.

Now we consider the minimum cost 
flow with maximal value $np$ in this network. Let its cost be  $w'\le w$.
This minimum cost flow corresponds to a schedule $\calX'$ in the following manner.
For each given $i$,
the amount of each job $j$ scheduled in interval $[t_i,t_{i+1})$ is equal to
the flow on the arc $(u_j, v_i)$, which is bounded by  the capacity
$\ell = t_{i+1}-t_i$ of this arc. The total processing time
$m\ell$ in this interval is assigned to jobs $1,2,\dots,n$, in this order,
processor by processor, and for each processor from left to right.
Since each $(u_j,v_i)$ has capacity $\ell$, a job will not be scheduled at
two processors at the same time. Also, since the capacity of the arc
between $v_i$ and the sink is $m\ell$, all jobs will be allocated
the required processing time.

Since all arcs have integer capacity,
the minimum cost flow can be assumed to be integer (see \cite{schrijver:combinatorial}.)
Therefore the resulting schedule $\calX'$ is integral.
It remains to show that its objective value is not larger than that of $\calX$.

For each job $j$,
let $C'_j$ be the completion time of $j$ in $\calX'$.
By the construction of the network, we have $C'_j\le \lceil C_j \rceil$. 
Moreover, for $w'$ jobs $j$ we have $C'_j = \lceil C_j \rceil > \lfloor C_j\rfloor$,
and for these jobs $C_j$ is not integer, while for all the other $n-w'$ jobs $j$
we have $C'_j \le \lfloor C_j \rfloor$. 
Setting $\tilde C=\sum_{j=1}^n \lfloor C_j \rfloor$, the cost of $\calX'$ is
\begin{eqnarray*}
\sum_{j=1}^{n} C'_{j} \;\le\; \tilde C + w' 
                \;\le\; \tilde C +w 
		 \;\le\; \tilde C + \sum_{j=1}^n (C_j- \lfloor C_j \rfloor )
		\;=\; \sum_{j=1}^{n} C_{j},
\end{eqnarray*}
completing the proof of the lemma.
\end{proof}


\paragraph{Busy schedules.}
We now show that, for the purpose of minimizing our objective
function, we can restrict our attention to schedules with some
additional properties.

We say that a schedule $\calX$ is \emph{busy} if it satisfies the
following condition: for any two times $s < t$ and a job $j$
such that $r_j \le s$, if $j\in \calX(t)$ and $|\calX(s)| < m$ (that is,
some machine is idle at $s$) then $j\in\calX(s)$ as well.  Any
optimal schedule is busy, for otherwise, if the above
condition is not satisfied, we can move a sufficiently small portion
$\epsilon > 0$ of $j$ from the last block where it is executed to the
interval $[s,s+\epsilon)$, obtaining a feasible schedule in which the
completion time of $j$ is reduced by $\epsilon$ and other completion
times do not change. Thus we only need to be concerned with
busy schedules.


\paragraph{Reductions.}
Let $\calX$ be a schedule, and $i,j$ be two jobs with $r_i<r_j$. Let
$T$ be the set of times where exactly one of the jobs $i,j$ is scheduled,
that is
$T= [\calX^{-1}(i)\setminus \calX^{-1}(j)] \cup [\calX^{-1}(j)\setminus \calX^{-1}(i)]$.
Also, let $t_0$ be a time point which divides $T$ into two parts of the same size,
that is $|T\cap[0,t_0)|=\frac12|T|$. 
Since $i$ and $j$ have equal processing time, the processing times
of each of $i$ and $j$ in $T$ is equal $\frac12|T|$.
The \emph{$(i,j)$-reduction} modifies the schedule by
executing $i$ in $|T\cap[0,t_0)|$ and $j$ in $T\cap [t_0,\infty)$.
If the reduction does not change $\calX$ (that is,
$i$ is executed in $|T\cap[0,t_0)|$ and $j$ in $T\cap [t_0,\infty)$)
then we say that \emph{$i$ and $j$ are in order}.  
 
We say that a schedule $\calX$ is \emph{irreducible} if it is
busy and all pairs of jobs are in order. It is not
difficult to see that $\calX$ is irreducible iff it satisfies
the following condition for any times $s < t$:
\begin{eqnarray}
        \max(\,\calX(s) - \calX(t)\,) &<& \min(\, \calX(t) - \calX(s)\,),
        \label{eqn: irreducible}
\end{eqnarray}
where the order refers to job indices, which we assumed to satisfy $i<j \Rightarrow r_{i}\leq r_{j}$.
Also, we use the convention
that $\max(\emptyset) = -\infty$ and $\min(\emptyset) = +\infty$,
so (\ref{eqn: irreducible}) holds whenever
$\calX(s)\subseteq \calX(t)$ or $\calX(t)\subseteq \calX(s)$.


\begin{lemma}
For any two jobs $i,j$ with $r_i < r_j$,
an $(i,j)$-reduction of $\calX$ does not increase the objective function,
and preserves the integrality and the
property of being busy.
\end{lemma}

\begin{proof}
In the $(i,j)$-reduction,
only the completion times of $i$ and $j$ might change. Let $C_i,C_j$ be the completion
times of $i$, $j$, before the reduction and $C'_i,C'_j$ after the reduction.
We have three cases.
If $C_i>C_j$ then $C'_j=C_i$ and $C'_i\le C_j$.
If $C_i=C_j$, then the completion times do not change.
If $C_i<C_j$, then $C'_j=C_j$ and $C'_i\le C_i$.
Thus in all three cases $C_i+C_j$ does not increase.
  
To justify the second part of the lemma, note that
the reduction does not change the cardinality of $\calX(t)$ at any time point $t$. Therefore the new schedule remains busy.
Further, if $\calX$ is integral then
$t_0$ can be assumed to be an integer, 
and the new schedule will then be integral as well.
\end{proof}


\begin{theorem}\label{lem: irreducible}
There is an optimal schedule $\calX$ that is irreducible.
\end{theorem}

\begin{proof}
Let $\calX$ be an optimal schedule that is busy and integral.
We define a potential function which decreases strictly when a
reduction of two jobs which are not in order is applied. From this we conclude
that after a finite number of reductions we must reach an irreducible schedule.

Define
$H(\calX) = (H_1(\calX), H_2(\calX),\ldots,H_n(\calX))$, where
$H_j(\calX) = \sum_{t\in\mathbb N, t \in \calX^{-1}(j)} t$ for
each job $j$. Given two different
schedules $\calX$, $\calY$, we say that $H(\calX)$ is
\emph{lexicographically smaller} than $H(\calY)$, if $H_i(\calX) <
H_i(\calY)$ for the smallest $i$ for which $H_i(\calX) \neq
H_i(\calY)$.

If two jobs $i<j$ are not in order, then the $(i,j)$-reduction
decreases $H_i(\calX)$, while $H_j(\calX)$ increases.
Therefore $H(\calX)$ decreases lexicographically. Any value
$H_j(\calX)$ is integer and bounded
by $0\le H_j(\calX)\le p(\max_i C_i)$, 
where $\max_i C_i$ is preserved by the reductions. From
this we can conclude that after a finite number of reductions
we obtain an irreducible optimal schedule.
\end{proof}


\medskip

We now give a characterization of irreducible schedules that will play
a major role in the construction of our linear program.

\begin{figure}[htb]
\centerline{\input{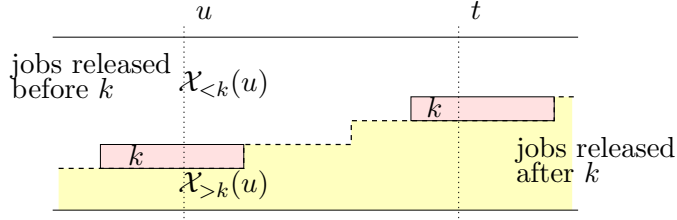}}
\caption{Structure of any irreducible schedule.}
\label{fig: normal structure}
\end{figure}

For a given job $j$ and a time $t$ we partition $\calX(t) -
\braced{j}$ into jobs released before and after 
$j$.  Formally, $\calXbef{j}(t) = \braced{i\in \calX(t): i<j}$ and
$\calXaft{j}(t) = \braced{i\in \calX(t): i>j}$.
The lemma below provides a characterization of
irreducible schedules. (See Figure~\ref{fig: normal structure}.)

\begin{lemma}\label{lem: order}
Let $\calX$ be an irreducible schedule.
Let $u,t$ be two time points and $k$ be a job such that $r_k \le u < t$.
 Then:

\noindent{\rm (a)}
If $k\in\calX(u) - \calX(t)$ then
$|\calXbef{k}(t)| \le |\calXbef{k}(u)|$.

\noindent{\rm (b)}
If $k\in\calX(t) - \calX(u)$ then
$|\calX(u)| = m$,  
$|\calXaft{k}(t)| \ge |\calXaft{k}(u)|$, and
$|\calXbef{k}(t)| < |\calXbef{k}(u)|$.

\noindent{\rm (c)}
If $k\in\calX(u) \cap \calX(t)$ then
$|\calXbef{k}(t)| \le |\calXbef{k}(u)|$ and
$|\calXaft{k}(t)| \ge |\calXaft{k}(u)|$.
\end{lemma}

\begin{proof}
Case  (a) $k\in\calX(u) - \calX(t)$:
We will show by contradiction that $\calXbef{k}(t) \subseteq \calXbef{k}(u)$, 
so suppose that there is a job
$j\in \calXbef{k}(t) - \calXbef{k}(u)$. 
Then $j \ge  
\min(\calXbef{k}(t) - \calXbef{k}(u)) = \min(\calX(t) - \calX(u))$. Also
$j<k$, and $k\le \max(\calX(u) - \calX(t))$.
This contradicts irreducibility by equation (\ref{eqn: irreducible}) and thus (a) follows.

Case (b) $k\in\calX(t) - \calX(u)$:
Since $k\not\in\calX(u)$ and $r_k \le u$, the assumption that
$\calX$ is busy implies that $|\calX(u)| = m$.

We must have $\calXaft{k}(u) \subseteq \calXaft{k}(t)$, for otherwise,
the existence of $k\in \calX(t) - \calX(u)$ and an $l \in
\calXaft{k}(u) - \calXaft{k}(t)$ would contradict irreducibility.  The
inequality $|\calXaft{k}(u)| \le |\calXaft{k}(t)|$ follows.  This, the
assumption of the case, and $|\calX(u)| = m$ imply $|\calXbef{k}(u)| >
|\calXbef{k}(t)|$.

Case (c) $k\in\calX(u) \cap \calX(t)$:
We only prove the first inequality, as the proof for the second
  one is very similar.  Towards contradiction, suppose
  $|\calXbef{k}(u)| < |\calXbef{k}(t)|$, and pick any
  $i\in\calXbef{k}(t) - \calXbef{k}(u)$.  Then $r_i\le r_k\le u$ and
  $i\in \calX(t)-\calX(u)$, and so the assumption that $\calX$ is
  busy implies $\calX(u) = m$.  This, in turn, implies that
  $|\calXaft{k}(u)| > |\calXaft{k}(t)|$, so we can choose $j\in
  \calXaft{k}(u) - \calXaft{k}(t)$.  But this means that $j > k > i$
  and $j\in \calX(u) - \calX(t)$, and the existence of such $i$ and
  $j$ contradicts irreducibility.
\end{proof}

\section{A Linear Program for $P|r_j;\pmtn;p_j=p|\sum C_j$}

\paragraph{Machine assignment.}
We now consider the actual job-machine assignment in an irreducible
schedule $\calX$.  As explained earlier, at every time $t$ we assign
the jobs in $\calX(t)$ to machines in order, that is job $j\in
\calX(t)$ is assigned to machine $1+|\calXbef{j}(t)|$.
Lemma~\ref{lem: order} implies that, for any fixed $j$, starting
at $t=r_j$ the value of
$|\calXbef{j}(t)|$ decreases monotonically with $t$. Therefore, with
machine assignments taken into account, $\calX$ will have the
structure illustrated in Figure~\ref{fig: normal structure}.

Call a schedule $\calX$ \emph{normal} if for each job $j$ and each
machine $q$, job $j$ is executed on $q$ in a single (possibly empty)
interval $[S_{j,q},C_{j,q})$, and
\begin{description}
\item{(1)} 
$C_{j,q}\le S_{j+1,q}$ for each machine $q$ and job $j<n$, and
\item{(2)}
$C_{j,q}\le S_{j,q-1}$ for each machine $q>1$ and job $j$.
\end{description}

\begin{figure}[htb]
\centerline{\epsfig{file=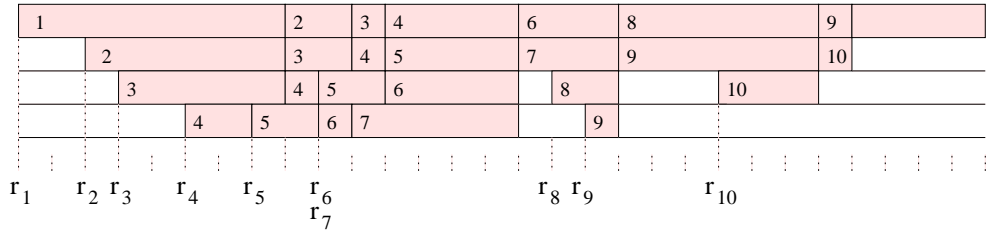,width=13cm}}
\caption{Example of a normal schedule. The processing
        time is $p=8$.}
\label{fig: normal example}
\end{figure}

By the earlier discussion, every irreducible schedule is normal
(although the reverse does not hold.)  An example of a normal (and
irreducible) schedule is shown in Figure~\ref{fig: normal example}.

\paragraph{Linear program.}
We are now ready to construct our linear program:
\begin{align}
{\mbox{\rm minimize }}   && \textstyle{\sum_{j=1}^n C_{j,1}}& 
                \label{eqn: linear program} \\
{\mbox{\rm subject to}} &&
           - S_{j,m}              \; &\le - r_j      && j = 1,\dots,n
                        \nonumber \\
        && \sum_q (C_{j,q} - S_{j,q})\; &= p         && j = 1,\dots,n
                          \nonumber \\
        && S_{j,q}   - C_{j,q}     \; &\le 0          && j = 1,\dots,n,\;  q = 1,\dots, m
                        \nonumber \\
        && C_{j,q} - S_{j,q-1}   \; &\le 0          && j = 1,\dots, n,\; q = 2,\dots,m
                        \nonumber \\
        && C_{j,q} -S_{j+1,q}  \; &\le 0          && j = 1,\dots,n-1,\;  q = 1,\dots,m
                        \nonumber
\end{align}


The correspondence between normal schedules and feasible solutions to
this linear program should be obvious. For any normal schedule, the
start times $S_{j,q}$ and completion times $C_{j,q}$ satisfy the
constraints of (\ref{eqn: linear program}).  And vice versa, for any
set of the numbers $S_{j,q}$, $C_{j,q}$ that satisfy the constraints
of (\ref{eqn: linear program}), we get a normal schedule by scheduling
any job $j$ in interval $[S_{j,q},C_{j,q})$ on each machine $q$.  Thus
we can identify normal schedules $\calX$ with feasible solutions of
(\ref{eqn: linear program}).  Note, however, that in $\calX$ a job $j$
could complete earlier than $C_{j,1}$ (this can happen when $C_{j,1} =
S_{j,1}$.) Thus the only remaining issue is whether the optimal normal
schedules correspond to optimal solutions of (\ref{eqn: linear
  program}).


\begin{theorem} 
The linear program above correctly computes an optimal schedule.
More specifically, $\min_\calX\sum_jC_j = \min\sum_jC_{j,1}$,
where on the left-hand side the minimum is over normal schedules
$\calX$ with $C_j$ representing the completion time of job $j$ in $\calX$,
and on the right-hand side we have the optimal solution of the
linear program  {\rm (\ref{eqn: linear program})}.
\end{theorem}

\begin{proof}
$(\le)$
  By the correspondence between normal schedules and feasible
  solutions of (\ref{eqn: linear program}), discussed before the
  theorem, we have $C_j\le C_{j,1}$ for all $j$, and thus the $\le$
  inequality is trivial.
  
$(\ge)$
To justify the other inequality, fix an optimal irreducible (and
thus also normal) schedule $\calX$. We want to show a feasible
solution to (\ref{eqn: linear program}) for which $C_j = C_{j,1}$
for all $j$.
 
 Fix a job $j$, and let
  $[S_{j,g},C_{j,g})$ be the last (that is, the one with minimum $g$)
  non-empty execution interval of $j$. So $C_j = C_{j,g}$ in $\calX$.
  Consider a block $[s,t)$ where
  $t=C_{j,g}$.  By Lemma~\ref{lem: order}, all jobs executed on
  machines $1,2,\dots,g-1$ in $[s,t)$ are numbered lower than $j$.
  Further, by the ordering of completion times, they are not executed
  after $t$. Thus these jobs must be completed at $t$ as well. Therefore we
  can set $[S_{j,h},C_{j,h})=[t,t)$, for all machines $1\le h<g$.
  This gives a normal schedule in which $C_j = C_{j,1}$.

Having done this for all jobs, we will get numbers $S_{j,q}$ and
$C_{j,q}$ that satisfy all constraints of the linear program,
and such that $C_j = C_{j,1}$ for all $j$.
\end{proof}

\section{Unary NP-Hardness of {\meanflowsched}}

In this section we prove that without the assumption on equal processing
times the problem is strongly (unary) NP-hard.

\begin{theorem} 
The problem {\meanflowsched} is strongly NP-hard, that is,
it is NP-hard even if the numbers on input are represented in 
the unary encoding.
\end{theorem}

\begin{proof}
The proof is by reduction from {\threepart}. In an
instance $\calI$ of {\threepart} we have $3n+1$ numbers
$x_1,x_2,\dots,x_{3n},y$ such that
$\sum_{i=1}^{3n} x_i = ny$ and
$\onefourth y < x_i < \onehalf y$ for each $i$.
We want to determine whether there is a 
partition  of $\braced{1,2,\dots,3n}$ into
$n$ sets $S_1,\dots,S_n$ such that
$\sum_{i\in S_k} x_i = y$ for all $k$.
(By the assumption about the numbers $x_i$, in any
such partition all sets $S_k$ have exactly three elements.)

Given an instance of {\threepart} above, we construct
an instance $\calJ$ of {\meanflowsched} as follows.
Define $A = 6ny$ and $B = 18n^2y^2$.
We let $m= n$, and $N=4n+An$. We create three types of
jobs:
\begin{description}
\item[x-jobs:]
For each $j = 1,\dots,3n$, we create job $j$ with
$r_j = 0$ and $p_j = Ax_j$.
\item[B-jobs:]
For each $j = 3n+1,\dots,4n$, we create job $j$ with
$r_j = Ay$ and $p_j = B$.
\item[1-jobs:]
For each $j = 4n+1,\dots,N$, we create job $j$ with $r_j = Ay+B$ and
$p_j = 1$.
\end{description}

If $\calI$ is represented in unary, its size is
$\Theta(ny)$. Then the size of $\calJ$ (in unary
encoding) is polynomial and the above transformation works
in polynomial time.

Thus now it is sufficient to prove the following claim:
$\calI$ has a 3-partition if and only if $\calJ$ 
has a schedule with $\sum_{j=1}^N C_j \le D$, where
$D = 3n Ay + n (Ay+B) + n\sum_{i=1}^A(Ay+B+i)$.

$(\Rightarrow)$
Let $S_1,\dots,S_n$ be a 3-partition of $\calI$.
For each $k$ we have $\sum_{j\in S_k} p_j = Ay$.
So on machine $k$ we schedule the following jobs:
first the three x-jobs $j\in S_k$, completing the
last one at $Ay$, then one B-job, followed by
$A$ 1-jobs. The objective value of this schedule is
\begin{eqnarray*}
\sum_{j=1}^N C_j &\le& 3n Ay + n (Ay+B)
         +  n\sum_{i=1}^A(Ay+B+i) 
                \;=\; D.
\end{eqnarray*}

$(\Leftarrow)$
Suppose now that $\calI$ does not have a 3-partition.
Consider any schedule $\calX$ of $\calJ$ on $n$ machines. 
We want to prove that the value of the objective
function for $\calX$ exceeds $D$.

We consider easy cases first.
If some x-job is completed after time $Ay+B$ then
\begin{eqnarray*}
\sum_{j=1}^N C_j &>& Ay+B +  n(Ay+B) 
                        + n\sum_{i=1}^A(Ay+B+i) 
                \;>\; D.
\end{eqnarray*}
If some B-job is completed after time $Ay+2B$ then
\begin{eqnarray*}
\sum_{j=1}^N C_j &>& (n-1)(Ay+B) + Ay+2B
                        + n\sum_{i=1}^A(Ay+B+i) 
                \;\ge\; D.
\end{eqnarray*}
Suppose now that each x-job is completed no later
than at time $Ay+B$ and each B-job is completed no
later than at time $Ay+2B$.
Since $\calI$ does not have a 3-partition, some machine
is idle for $A$ time units in the interval $[0,Ay]$.
Therefore some B-job will complete between $Ay+B+A$ and
$Ay+2B$, keeping one machine busy in the interval
$[Ay+B,Ay+B+A]$. This implies that there are at
least $A$ 1-jobs that will be scheduled at time
$Ay+A+B$ or later. Thus
\begin{eqnarray*}
\sum_{j=1}^N C_j &>& (n-1)(Ay+B) + Ay+A+ B
                        + (n-1)\sum_{i=1}^A(Ay+B+i) 
                        + A(Ay+A+B)
                \;>\; D.
\end{eqnarray*}
Summarizing, in all cases the objective value exceeds $D$,
completing the proof of the claim.
\end{proof}

\section{Final Remarks} \label{sec:more}

We proved that the scheduling problem $P|r_j,\pmtn,p_j=p|\sum C_j$
can be reduced to solving a linear program with
$O(mn)$ variables and constraints. This leads to a simple polynomial
time algorithm. Since we can assume that $m\le n$ (otherwise the problem
is trivial) the running time can be also expressed as a
polynomial of $n$ only.

We showed that there is an optimal schedule with 
$O(mn)$ preemptions.  We do not know whether
this bound is asymptotically tight.  It is quite possible that
there exist optimal schedules in which the number of preemptions is
$O(n)$, independent of $m$. If this is true, this could lead to
even more efficient, combinatorial (not dependent on linear
programming) algorithms for this problem. Such
improvement, however, would require a deeper study of the structural
properties of optimal schedules.  Since we use only the existence of
normal optimal schedules, rather than irreducible schedules, we feel
that the problem has more structure to be exploited.

An interesting related open question is $P | r_j, p_j=p  | \sum U_j$,
where the objective is to find a maximal subset of jobs which can be scheduled on time.
It has been shown in \cite{kravchenko:open-shop} 
that the corresponding preemptive version,
$P | r_j, p_j=p, \mbox{pmtn}  | \sum U_j$, is unary NP-hard.

We implemented the complete algorithm (converting the instance to a
linear program and solving this linear program). It is accessible at
Christoph D\"urr's webpage.

\bibliographystyle{plain}
\bibliography{sched}

\appendix

\section{Proof of Theorem~\ref{lem: irreducible} for Arbitrary Real Numbers}
\label{sec: irreducible general case}

In Section~\ref{sec: preemptions} we proved that there is always an
optimal irreducible schedule, under the assumption that the processing
time $p$ and the release times $r_1,\dots,r_n$ are integral. We now show
that this is true in the more general scenario when all these numbers
are arbitrary reals.

We first extend the definition of the potential function. For any
job $j$, let
$\barH_j(\calX) = \onehalf \int_{\calX^{-1}(j)} t \, \mathrm{d}t$,
where the integral is taken over the support of $j$.
(We remark that this is a standard value in scheduling, even though
the factor $\onehalf$ is irrelevant for this paper.)
The generalized potential function is
$\barH(\calX) = (\barH_1(\calX),\ldots,\barH_n(\calX))$.

The following properties remain true:
(i) a reduction does not increase the objective
function and preserves the property of a schedule being busy,
(ii) $\barH(\calX)$ strictly decreases  (lexicographically)
after each reduction.

The major difficulty that we need to overcome is that the set of
schedules is not closed as a topological space, so there could be a
sequence of schedules with decreasing values of $\barH(\calX)$ whose
limit is not a legal schedule. The idea of the proof is to reduce
the problem to minimizing $\barH(\calX)$ over a compact subset of
schedules.


\begin{lemma}
Even if $p,r_1,\dots,r_n$ are arbitrary real numbers,
there is an irreducible optimal schedule.
\end{lemma}

\begin{proof}
  Define a \emph{block} of a schedule $\calX$ to be a
  maximal time interval $[u,t)$ such that $(u,t)$ does not contain any
  release times and $\calX(s)$ is constant for $s\in[u,t)$.
  
  For convenience, let $r_{n+1}$ be any upper bound on the last
  completion time of any optimal schedule, say $r_{n+1} = r_n+ np$.
  Thus all jobs are executed between $r_1$ and $r_{n+1}$.  Each
  interval $[r_i,r_{i+1})$, for $i=1,\dots,n$ is called a
  \emph{segment}.  By condition (s3), each segment is a disjoint union
  of a finite number of blocks of $\calX$.  Also, for each job $j$, we
  have $C_j = t$ for the last non-empty block $[s,t)$ whose profile
  contains $j$.
  
  A schedule $\calX$ is called \emph{tidy} if all jobs are completed
  no later than at $r_{n+1}$ and, for any segment $[r_i,r_{i+1})$, the
  profiles $\calX(t)$, for $t\in [r_i,r_{i+1})$, are lexicographically
  ordered from left to right. More precisely, this means that, for any
  $r_i \le s < t < r_{i+1}$, we have
\begin{eqnarray*}
        \min(\,\calX(s) - \calX(t)\,) &\le& \min(\, \calX(t) - \calX(s)\,).
\end{eqnarray*}
One useful property of tidy schedules is that its total number of
blocks (including the empty ones) is $N = nM$, where $M = \sum_{l =
  0}^m\binom{n}{l}$, because $n$ is the number of intervals $[r_{i},r_{i+1})$ and $M$ is the number of lexicographically ordered blocks in each interval.  
  From now on we identify any tidy schedule
$\calX$ with the vector $\calX\in \reals^N$ whose $b$-th coordinate
$x_b$ represents the length of the $b$-th block in $\calX$.

In fact, the set $\bfT$ of tidy schedules is a (compact) convex
polyhedron in $\reals^N$, for we can describe $\bfT$ with a set of
linear inequalities that express the following constraints:
\begin{itemize}
\item Each job $j$ is not executed before $r_j$,
\item Each job $j$ is executed for time $p$.
\end{itemize}
For example, the second constraint can be written as $\sum_b x_b = p$,
where the sum is over all blocks $b$ whose profile contains $j$.


\begin{claim} \label{cla: completion ordering}
Any schedule $\calX$ can be transformed into a schedule
with completion times ordered as $C_1\le \ldots\le C_n$, without increasing the
objective function value.
\end{claim}

\begin{proof}
Note that after an $(i,j)$-reduction we have $C_i\le C_j$, and no other completion
time is changed. Therefore after reducing job $1$ successively with the jobs
$2,\ldots,n$, $C_1$ is the smallest completion time.  Then after reducing job
$2$ with the jobs $3,\ldots,n$ we have $C_1\le C_2 \le \min\{C_3,\ldots,C_n\}$.
Continuing this process will eventually end with ordered completion times.
\end{proof}


\begin{claim}\label{cla: tidy schedules}
  Let $\calX$ be a schedule in which completion times are ordered and
  upper bounded by $r_{n+1}$.  Then $\calX$ can be converted into a
  tidy schedule $\calX'$ such that

\noindent {\rm (a)} $C'_j \le C_j$ for all $j$ (where $C_j$ and $C'_j$
        are the completion times of $j$ in $\calX$ and $\calX'$,
        respectively.) 

\noindent {\rm (b)} $\barH(\calX')$ is equal to or lexicographically
         smaller than $\barH(\calX)$.
\end{claim}

\begin{proof}
Indeed, suppose that $\calX$ has two consecutive blocks $A = [u,s)$,
$B = [s,t)$ where $r_i\le u < s < t \le r_{i+1}$, and the profile
$\calX(u)$ of $A$ is larger (lexicographically) than the profile
$\calX(s)$ of $B$.  Exchange $A$ and $B$, and denote by $\calX'$ the
resulting schedule.  Let $j = \min(\calX(s)-\calX(u)) <
\min(\calX(u)-\calX(s))$.  Since $C_j \ge t$, all jobs in
$\calX(u)-\calX(s)$ are also completed not earlier than at $t$. So
this exchange does not increase any completion times.  We have
$\barH_j(\calX') < \barH_j(\calX)$ and $\barH_i(\calX') = \barH_i(\calX)$ for $i<j$.
Thus $\barH(\calX')$ is lexicographically smaller than $\barH(\calX)$.  By
repeating this process, we eventually convert $\calX$ into a tidy
schedule that satisfies the claim.
\end{proof}

We now continue the proof of the lemma.  Fix some optimal schedule
$\calX^\ast$.  Let $C^\ast_j$ denote the completion time of a job $j$
in $\calX^\ast$.  From Claims~\ref{cla: completion ordering} and
\ref{cla: tidy schedules}, we can assume that $\calX^\ast$ is tidy and
$C^\ast_1 \le C^\ast_2 \le \ldots \le C^\ast_n$.  (For the peace of
mind, it is worth noting that Claim~\ref{cla: tidy schedules} implies
that $\calX^\ast$ is well defined, for it reduces the problem to
minimizing $\sum_j C_j$ over a compact subset $\bfT$ of $\reals^N$.)
 
Consider a class $\bfT_0 \subseteq \bfT$ of tidy schedules $\calX$
such that each job $j$ in $\calX$ is completed not later than
at time $C^\ast_j$. Since $\calX^\ast\in\bfT_0$, the set $\bfT_0$ is not
empty.  Similarly as $\bfT$, $\bfT_0$ is a (compact) convex
polyhedron.  Indeed, we obtain $\bfT_0$ by using the same constraints
as for $\bfT$ and adding the constraints that each job $j$ is
completed not later than at $C^\ast_j$.  To express this constraint,
if in $\calX^\ast$ the completion time $C^\ast_j$ of $j$ is at the end
of the $a$-th block in the segment $[r_i,r_{i+1})$, then for each $b >
(i-1)M+a$ such that $j$ is in the profile of the $b$-th block, we
would have a constraint $x_b = 0$.  Note that these constraints do not
explicitly force $j$ to end \emph{exactly} at $C^\ast_j$, but the
optimality of $\calX^\ast$ guarantees that it will have to.

Now we show that there exists a schedule $\calY^\ast \in\bfT_0$ for
which $\barH(\calY^\ast)$ is lexicographically minimum.  First, as we
explained earlier, $\bfT_0$ is a compact convex polyhedron.  Let
$\bfT_1\subseteq \bfT_0$ be the set of $\calX$ for which $\barH_1(\calX)$
is minimized.  $\barH_1$ is a continuous quadratic function over $\bfT_0$,
and thus $\bfT_1$ is also a non-empty compact set.  Continuing this
process, we construct sets $\bfT_2, \ldots, \bfT_n$, and we choose
$\calY^\ast$ arbitrarily from $\bfT_n$.
\end{proof}

\end{document}